\documentclass[pra,twocolumn,showpacs,amsmath,amssymb,superscriptaddress]{revtex4-1}
\usepackage{graphicx,bm,color,mathptmx,hyperref} 
\begin{document}

\title{QED with a parabolic mirror}
\author{G. Alber}
\affiliation{Institut f\"{u}r Angewandte Physik, 
Technische Universit\"{a}t Darmstadt, 64289 Darmstadt, Germany}

\author{J. Z. Bern\'{a}d}
\affiliation{Institut f\"{u}r Angewandte Physik, 
Technische Universit\"{a}t Darmstadt, 64289 Darmstadt, Germany}

\author{M. Stobi\'nska}
\affiliation{Institute of Theoretical Physics and Astrophysics, 
University of Gda\'nsk, ul. Wita Stwosza 57, 80-952 Gda\'nsk, Poland}
\affiliation{Institute of Physics, Polish Academy of Sciences, 
Al. Lotnik\'ow 32/46, 02-668 Warsaw, Poland}

\author{L.~L.~S\'{a}nchez-Soto} 
\affiliation{Max-Planck-Institut f\"ur die Physik des Lichts,
  G\"{u}nther-Scharowsky-Stra{\ss}e 1, Bau 24, 91058 Erlangen,
  Germany} 
\affiliation{Department f\"{u}r Physik, Universit\"{a}t Erlangen-N\"{u}rnberg,
Staudtstra{\ss}e 7, Bau 2, 91058 Erlangen, Germany}
\affiliation{Departamento de \'Optica, Facultad de Fisica, Universidad
  Complutense, 28040 Madrid, Spain}

\author{G. Leuchs}
\affiliation{Max-Planck-Institut f\"ur die Physik des Lichts,
  G\"{u}nther-Scharowsky-Stra{\ss}e 1, Bau 24, 91058 Erlangen,
  Germany} 
\affiliation{Department f\"{u}r Physik, Universit\"{a}t Erlangen-N\"{u}rnberg,
Staudtstra{\ss}e 7, Bau 2, 91058 Erlangen, Germany}

\date{\today}

\begin{abstract}
  We investigate the quantum electrodynamics of a single
  two-level atom located at the focus of a parabolic cavity.  We first
  work out the modifications of the spontaneous emission induced by
  the presence of this boundary in the optical regime, where the
  dipole and the rotating-wave approximations apply.  Furthermore, the
  single-photon state that leaves the cavity asymptotically is
  determined.  The corresponding time-reversed single-photon quantum
  state is capable of exciting the atom in this extreme multimode
  scenario with near-unit probability. Using semiclassical methods, we
  derive a photon-path representation for the relevant transition
  amplitudes and show that it constitutes a satisfactory approximation
  for a wide range of wavelengths.
\end{abstract}


\maketitle

\section{Introduction}

The physics of strong light-matter coupling has been attracting a
great deal of attention over the last few years. Since the early
eighties, single atoms have been coupled to optical and microwave
cavities, leading to fundamental demonstrations of cavity
quantum electrodynamics
(QED)~\cite{Berman:1994cd,Walther:2006qc,Haroche:2006pm}.  More
recently, impressive developments in circuit QED, involving
superconducting qubits coupled to microwave
cavities~\cite{Wallraff:2004hj}, atom
chips~\cite{Colombe:2007qc}, and chip-based
microresonators~\cite{Aoki:2006jk} have opened the door
to the ultra-strong coupling regime, holding the promise to exploit
light-matter interaction at the single-photon level in scalable
architectures. This is of pivotal significance€ for future
applications of quantum technologies.

Thus far, in typical cavity QED configurations, the atomic properties
are considerably changed because the cavity modifies the
electromagnetic mode structure as a consequence of the
boundaries~\cite{Boundaries}.  Actually, the radiating atom can excite
only one or a few radiation modes~\cite{radmodes}.  In these cases, we
can observe spontaneous emission enhancement or inhibition into the
modes that are resonant or non-resonant with the cavity, respectively.

The extreme opposite regime of free-space QED, where a continuum of
modes are available, has also received notable
recognition~\cite{contmodes},
motivated by the hope of finding simpler solutions for quantum
communication over large distances.  In these circumstances, it is
essential to increase the strength of the light-matter interaction:
strongly focused light improves the
coupling~\cite{focusing} and matching the incoming
field with the spatial atomic radiation mode improves 
focusing~\cite{matching}. Moreover, tailoring the
polarization pattern can be significant for achieving near perfect 
coupling~\cite{tailorpol}.

An intriguing intermediate instance between the single-mode and the
continuum limit is the case of a large
cavity~\cite{Alber:1992qr,Daul:2005se,Viviescas:2003pi}, in which an
atom couples to a large but not continuous number of modes.  A
half-cavity, i.e., a cavity with one mirror, constitutes a good example
of such a situation~\cite{Dorner:2002pz,StAl}. It has been verified
experimentally that also in this regime one can witness a change of
the density of field modes near an atom which manifests itself in a
modified spontaneous photon emission rate~\cite{Eschner:2001aj,StSoLe}.

A parabolic cavity is a remarkable example of a half-cavity. The
parabolic shape ensures that light entering parallel to the symmetry
axis couples to an atom located at its focus in a particularly
efficient way, the light impinging on the atom from all
directions~\cite{Bokor:2008pi}. Conversely, such a parabola collects
the light emitted by an atom in the spontaneous decay in all
directions.

In a classical ray picture, valid for focal lengths of the parabola
large in comparison with the relevant wavelengths of the radiation,
only the small fraction of radiation emitted by the atom along the
symmetry axis in the direction of the vertex of the parabola is
back-reflected towards the atom. Thus, it might seem that the atom
scarcely feels the presence of boundaries.  However, this picture is
largely oversimplified: if the focal length of the parabola becomes
comparable to the relevant radiative wavelengths, diffraction effects
become important and significant modifications of the spontaneous
emission can be expected.  Besides, the atom is not a point, but it
scatters photons within a region whose linear extension is of the
order of the wavelength. Thus, both diffraction and resonant photon
scattering by the atom are expected to modify the simple
short-wavelength picture substantially.

Prompted by the current interest in radiative effects in half-open
cavities, we look here into the QED of a two-level atom located at the
focus of a parabolic mirror. First, we find the vector field modes
that can couple efficiently to the atom in the dipole
approximation. In terms of them, we study the ensuing modifications of
the spontaneous emission as well as the quantum statistical space-time
properties of the generated photon.

For that purpose, we develop a semiclassical path representation of
probability amplitudes that interpret them as sums of contributions
associated with different photon paths inside the parabolic cavity.
Exploiting in a systematic way the separability of the Helmholtz
equation in parabolic coordinates, such a representation provides an
adequate quantitative description of the spontaneous emission, not
only in the short-wavelength limit, but also in the regime of
wavelengths comparable or even smaller than the focal length of the
parabolic cavity.

The plan of this paper is as follows. In Sec.~\ref{Basics} the basic
model and the approximations involved are summarized.  The dynamics of
the spontaneous decay process is described in Sec.~\ref{Dynamics}: the
decay rate and its dependence on the focal length of the parabola is
discussed in Sec.~\ref{Rate}, while the subsequent subsection explores
the conditions under which the spontaneous emission can be described
by an exponential decay and the modifications that occur due to the
presence of the cavity.  Finally, in Sec.~\ref{OnePhoton}
characteristic properties of the spontaneously generated one-photon
quantum state are discussed.

\section{Setting the model 
\label{Basics}}

\subsection{Atom-field interaction in a parabolic mirror}

\begin{figure}
\includegraphics[width=0.80\columnwidth]{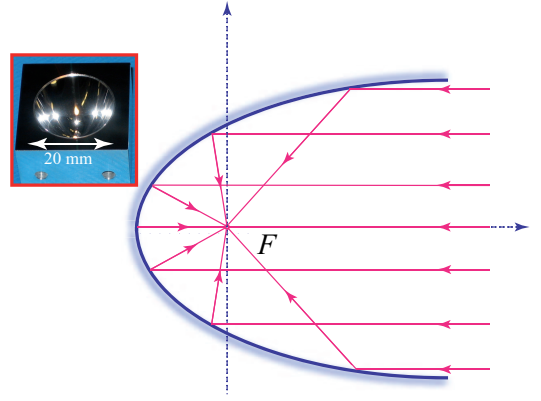}
  \caption{(Color online) Schematic representation of the parabolic cavity: the
    two-level atom is situated at the focus $F$ with $\mathbf{x}_0 =
    0$.  In parabolic coordinates the boundary of this cavity is
   given by the equation $\eta = 2f$. We also  include a picture of
   the real mirror in the Erlangen experiment, has a focal length of
   $f = 2.1$ mm  with a front opening of 20 mm in diameter, resulting
   in a  depth of 11.9 mm.}
  \label{Fig:parabola}
\end{figure}

We consider an atom situated at the focus $\mathbf{x}_{0}$ of an
axially symmetric parabolic cavity, as sketched in
Fig.~\ref{Fig:parabola}.  We take the atom initially prepared in an
excited electronic state, say $|e\rangle$, that decays by an allowed
dipole transition to the electronic ground state $| g\rangle$.  In the
Schr\"odinger picture, we can model this atom by a two-level system
with the Hamiltonian
\begin{equation}
  \hat{H}_A = 
  E_{e}~|e\rangle \langle e| + 
  E_{g}~|g\rangle \langle g| \, . 
\end{equation}

The free evolution of the quantized radiation field inside this cavity
is described by the standard Hamiltonian
\begin{equation}
\label{HF}
  \hat{H}_F =  \sum_{n} \int d\omega~\hbar
  \omega~\hat{a}^{\dagger}_{\omega,n}\hat{a}_{\omega,n} \, ,
\end{equation}
which has to include all the modes which couple quasi-resonantly to
the atom.  These modes (whose explicit form will be determined in the
next subsection) are labeled by their continuous frequencies $\omega$
and by a discrete parameter $n$ that incorporates the boundary
effects. In Eq.~(\ref{HF}), $\hat{a}^{\dagger}_{\omega,n}$ and
$\hat{a}_{\omega,n}$ are the creation and destruction operators of the
corresponding modes, respectively.

We recall that, in the Schr\"odinger picture, the operators of the
electric field $\hat{\mathbf{E}} (\mathbf{x})$ and of the magnetic
field $\hat{\mathbf{B}} ( \mathbf{x})$ of these modes are given by 
\begin{eqnarray}
  \hat{\mathbf{E}} (\mathbf{x}) & = &
  i~\sqrt{\frac{\hbar \omega}{2\epsilon_0}} \,
 \sum_{n} \int_0^{\infty} d\omega~[ \mathbf{g}_{\omega,n} (\mathbf{x}) \,
 \hat{a}_{\omega,n} -   {\rm H. \, c.} ] \, , 
  \nonumber \\
  & & \\
  \hat{\mathbf{B}} (\mathbf{x}) &=& 
 \sqrt{\frac{\hbar \omega}{2\epsilon_0}} \, 
  \sum_{n} \int_0^{\infty }d\omega~[ \nabla \times 
  \mathbf{g}_{\omega,n} ( \mathbf{x}) \, \hat{a}_{\omega,n} +
   {\rm H. \, c}] \, ,
  \nonumber
\end{eqnarray}
with ${\rm H.c.}$ denoting the Hermitian conjugate operators.  The
orthonormal mode functions $\mathbf{g}_{\omega,n}(\mathbf{x})$ fulfill
the transversality condition $ \nabla \cdot \mathbf{g}_{\omega,n}
({\bf x}) = 0$ and they are solutions of the vectorial Helmholtz
equation
\begin{equation}
  \left(\Delta + \frac{\omega^2}{c^2} \right) 
  \mathbf{g}_{\omega,n}(\mathbf{ x}) = 0 \, ,
  \label{Helmholtz}
\end{equation}
with $c$ the speed of light in vacuum.  This equation has to be
interpreted in the sense that it applies to each Cartesian component
$\mathbf{e}_{i} \cdot \mathbf{g}_{\omega,n} (\mathbf{x})$ of the mode
function separately. The orthonormality condition reads
\begin{eqnarray}
  \int_{{\mathbb R}^3} d^3 \mathbf{x}~\mathbf{g}_{\omega,n}^{\ast }(\mathbf{x})
  \cdot   \mathbf{g}_{\omega^{\prime},n^{\prime}} (\mathbf{x}) = 
  \delta_{n n^{\prime}} \delta(\omega - \omega^{\prime}) \, .
  \label{orthonormal}
\end{eqnarray}

%%%%%%%%%%%%%%%%%%%%%%%%%%%%%%%%%%%%%%%%%%%%%%%%%
%%%%%%%%%%%%%%%%%%%%%%%%%%%%%%%%%%%%%%%%%%%%%%%%

In the dipole approximation, the atom-field interaction is described
by $-\hat{\mathbf{d}} \cdot \hat{\mathbf{E}} ( \mathbf{x}_0)$, with
$\hat{\mathbf{d}}$ being the atomic dipole operator. In the optical
range, where the rotating-wave approximation is valid, this coupling
reduces to
\begin{equation}
  \hat{H}_{AF}  =  - i~  \sqrt{\frac{\hbar\omega}{2\epsilon_0}} \, 
  \sum_{n}\int d\omega~[  \mathbf{d} \cdot \mathbf{g}_{\omega,n} ( \mathbf{x}_0 ) \,
  \hat{a}_{\omega,n} \, |e\rangle \langle g| + {\rm H. c.} ] \, , 
  \label{Hint}
\end{equation}
where $\mathbf{d}= \langle e|\hat{\mathbf{d}} |g\rangle$ is the
atomic-dipole matrix element between the excited state $|e\rangle$ and
the ground state $|g\rangle$.

To assess the dynamics of the spontaneous emission one has to solve
the time-dependent Schr\"odinger equation with the Hamiltonian
\begin{equation}
  \hat{H} = \hat{H}_A + \hat{H}_F + \hat{H}_{AF} 
  \label{Hamiltonian}
\end{equation}
and the initial condition that at time $t_0$ the state of the
atom-field system is
\begin{equation}
  |\psi(t_0)\rangle =  |e\rangle \otimes |0\rangle \, ,
  \label{initialcondition}
\end{equation}
$|0\rangle$ being the ground (vacuum) state of the free
electromagnetic field.

\subsection{Normal modes and quantization
 \label{modefunctions}}

We assume the two-level atom located at the focus of the parabolic
cavity (which we take as $\mathbf{x}_0 = 0$), with its transition dipole
matrix element $\mathbf{d}$ oriented along the symmetry axis of the
parabola, i.e.,  $\mathbf{d} = d\mathbf{e}_3$.  This is the case in the
experimental setup in our laboratory~\cite{Maiwald:2012ex}. As a
result, this atom couples only to those modes whose electric field at
the focus is oriented along the symmetry axis.  For distances close to
the atom, i.e. $|\mathbf{x} - \mathbf{x}_0|\omega/c \ll 1$, these mode
functions are not modified by the boundary conditions and are of the
same form as in free space, i.e.,
\begin{equation} 
\mathbf{g}_{\omega,n}(\mathbf{x})  \rightarrow \nabla
  \times C \rho \mathbf{e}_{\varphi} = 2 C \mathbf{e}_{3} \, ,
  \label{small}
\end{equation}
where $C$ is a normalization constant and we have used cylindrical
coordinates $(\rho, \varphi, z)$. The vector $\mathbf{e}_{\varphi} =
\cos\varphi \mathbf{e}_2 - \sin\varphi \mathbf{e}_1$ denotes the unit
tangent vector in the angular direction $\varphi$.

Far away from the atom, however, these modes are altered by the
presence of the parabola. Transversality can be ensured by imposing
\begin{equation}
  \mathbf{g}_{\omega,n}(\mathbf{x}) = \nabla \times  
  \mathbf{G}_{\omega,n} (\mathbf{x}) \, ,
\end{equation}
so that $\mathbf{G}(\mathbf{x}) = f(x,y) \mathbf{e}_{\varphi}$ is also
a solution of the Helmholtz equation.

The separability of that equation in parabolic coordinates, suggests
the \emph{ansatz} (see the Appendix)
\begin{equation}
  \mathbf{G}_{\omega,n} (\mathbf{x}) = 
 \frac{1} {\sqrt{2 \pi \omega \mathcal{N}_{\omega,n}}} 
  \frac{\chi_{\omega,n} (\xi)}{\sqrt{\xi}} 
  \frac{\chi_{\omega,n} (\eta)} {\sqrt{\eta}} 
  \mathbf{e}_{\varphi} \, ,
  \label{modeansatz}
\end{equation}
provided the regular functions $\chi_{\omega,n}(\xi)$ and
$\chi_{\omega,n}(\eta)$ fulfill 
\begin{eqnarray}
  \left( \frac{d^2 }{d \xi^2} + 
    \frac{\omega^2}{4c^2} - 
    \frac{\alpha}{\xi}\right) \chi_{\omega,n}(\xi) = 0 \, , 
  \nonumber\\
  & &
  \label{Coulomb}\\
  \left(\frac{d^2 }{d \eta^2} + 
    \frac{\omega^2}{4c^2} + 
    \frac{\alpha}{\eta}\right) \chi_{\omega,n}(\eta) = 0 \, . 
  \nonumber  
\end{eqnarray}
For each frequency $\omega$, the possible values $\alpha_n$ of the
separation constant $-\infty < \alpha < \infty$ have to be determined
by the boundary conditions at the surface of the parabola.  The
frequency-normalized solutions of Eqs.~(\ref{Coulomb}) can be
expressed in terms of the Coulomb functions~\cite{AS}
\begin{eqnarray}
  \chi_{\omega,n}(\xi) & = & \sqrt{\frac{4}{\pi k}}
  F_{L=0}(\alpha_n/k, k \xi/2) \, , 
 \nonumber \\
  & &  \\
  \chi_{\omega,n}(\eta) & = & \sqrt{\frac{4}{\pi k}}
  F_{L=0}(-\alpha_n/k, k \eta/2) \, , 
  \nonumber 
  \label{separation}
\end{eqnarray}
with $k:= \omega/c$. In lieu of this procedure, one could also find $
\mathbf{G}_{\omega,n} (\mathbf{x}) $ from the general solution of the
scalar Helmholtz equation in parabolic coordinates worked out in
Ref.~\cite{Boyer:1976vt} by imposing the Coulomb gauge condition and
the appropriate boundary condition at a later stage. However, in
general such an approach may lead to formidable mathematical problems.

In these coordinates, the parabola of focal length $f$ is given by the
equation $\eta = 2f$, so the normalization factor in
Eq.~(\ref{modeansatz}) involves the quantity
\begin{equation}
  \mathcal{N}_{\omega,n} = \int_0^{2f}
  d\eta~\frac{\chi_{\omega,n}^2(\eta)}{\eta} \, ,
  \label{normal1}
\end{equation}
and has been chosen in such a way that the associated mode function
$\mathbf{g}_{\omega,n}(\mathbf{x})$ satisfies the orthonormalization
condition~(\ref{orthonormal}).

The quantization of the separation constant $\alpha$ is determined by
the boundary conditions at the surface of the parabola.  To simplify
the details as much as possible, we take a perfect metallic mirror
with no losses and reflection coefficient $r = -1$ (we neglect any
dependence of $r$ on the angle of incidence or on the
wavelength). This corresponds to imposing the tangential components of
the electric field and the normal component of the magnetic field to
vanish, which is warranted whenever
\begin{equation}
  \left . \frac{d\chi_{\omega,n}}{d\eta} \right |_{\eta = 2f} = 0 \, ,
  \label{quantization1}
\end{equation}
wherefrom the permitted values $\alpha_n$ can be determined for each
possible frequency $\omega$ and for $n \in {\mathbb N}$.

In the semiclassical limit, $\chi_{\omega,n}^2(\eta)$ is a rapidly
oscillating function over the range of integration in
Eq.~(\ref{normal1}). The quantization condition can be encoded in an
eikonal function $S(\omega, \alpha) = \pi [ n(\omega, \alpha) + 1/2]$
and the normalization factor fulfills the characteristic relation
\begin{equation}
  \mathcal{N}_{\omega,n} = 2\frac{\partial n} 
  {\partial \alpha}(\omega,\alpha_n) \, ,
  \label{normal2}
\end{equation}
which establishes an important relation between the separation
constant $\alpha$ and the quantization number $n$. We recall that for
the setup in our laboratory in Erlangen, where $f = 2.1$ mm and a
wavelength of 369 nm is used, this semiclassical limit is well satisfied.

\section{QED effects at the focus of a parabolic mirror
 \label{Dynamics}}

In this Section, we first discuss the dependence of the spontaneous
decay rate $\Gamma(\omega_0)$ on the focal length $f$ of the parabola
in the framework of a time-dependent perturbation theory.

The second subsection is devoted to an investigation of the dynamics
of the spontaneous emission.  In particular, we demonstrate that for
moderate focal lengths the spontaneous decay is exponential, whereas,
if the focal length is large enough so that subsequent reflections of
the photon at the parabola can be distinguished in space-time, this
exponential decay is appreciably modified.

For the quantitative analysis of this latter phenomenon a photonic
semiclassical path representation is developed. In the spirit of the
path integral approach~\cite{Feynman}, it resolves the probability
amplitudes of interest into contributions corresponding to all
possible photon paths and their multiple reflections at the parabolic
mirror.  This picture also sheds light onto the validity of the
pole approximation~\cite{poleapproximation}.

\subsection{The spontaneous decay rate}
\label{Rate}

The spontaneous decay rate characterizes the basic aspects of the
spontaneous emission of a photon of frequency $\omega_0 = (E_e -
E_g)/\hbar$. In the dipole approximation and in the lowest order of
time-dependent perturbation theory~\cite{GoldenRule} it is given by
\begin{equation}
  \Gamma (\omega_0) =  \frac{2\pi}{\hbar^2} 
  \sqrt{\frac{\hbar \omega_0}{2\epsilon_0}}
  \sum_{n} \left |   \mathbf{d} \cdot   
    \mathbf{g}_{\omega_0,n} ( \mathbf{x}_{0} ) \right | ^2 \, . 
\end{equation}
Using the mode functions $\mathbf{g}_{\omega,n}(\mathbf{x})$ we obtain
for our case
\begin{equation}
  \Gamma (\omega_0) = 
  \Gamma_s (\omega_0)  \frac{6}{\pi k_0} 
  \sum_n \frac{1}{\mathcal{N}_{\omega_0,n}}
  \left ( \frac{\pi\alpha_n(\omega_0)/ k_0}
    {\sinh[ \pi\alpha_n (\omega_0)/k_0 ]} \right )^2 \, ,
  \label{Fermiexact}
\end{equation}
with $k_0 = \omega_0/c$ and
\begin{equation}
  \Gamma_s (\omega_0) =  
  \frac{ | \mathbf{d} |^{2} \omega_0^3}{3\pi \epsilon_0\hbar c^3}
  \label{Gammafree}
\end{equation}
being the free-space spontaneous decay
rate. Equation~(\ref{Fermiexact}) conveys all the modifications in the
spontaneous emission brought about by the parabolic mirror. As we can
see, it involves a sum over all the quantized separation constants
$\alpha_n(\omega_0)$, which makes its explicit evaluation difficult,
except if only a few values of the separation constants
$\alpha_n(\omega_0)$ contribute dominantly to the summation.

Alternatively, we can rewrite Eq.~(\ref{Fermiexact}) making use of
the semiclassical relation~(\ref{normal2}) and of the Poisson
summation formula~\cite{PoissonSummation} as
\begin{equation}
  \Gamma (\omega_0)  = \Gamma_s(\omega_0)
  \sum_{M=-\infty}^{\infty} \frac{3}{\pi^2} 
  \int_{-\infty}^{\infty}dx~\frac{x^2}
  {\sinh^2x} 
 \exp[ i 2\pi M n(\omega_0,x) ] \, , 
  \nonumber \\
  \label{Poisson}
\end{equation}
where $x:= \pi \alpha_n (\omega_0)/k_0$. This form is particularly
convenient if the exponential functions involved in the integration
over $x$ are rapidly oscillating functions.  In these cases, the
dominant contribution of Eq.~(\ref{Poisson}) comes from the term with
$M=0$ with smaller contributions originating from the terms with
$M\neq 0$. As a matter of fact, the contribution of $M=0$ yields
precisely 1, and consequently, if the contributions to
Eq.~(\ref{Poisson}) resulting from $M \neq 0$ are neglected, the
spontaneous decay rate $\Gamma(\omega_0)$ reduces to its value in free
space $\Gamma_s(\omega_0)$.

\begin{figure}[t]
  \begin{center}
    \includegraphics[width=0.85\columnwidth]{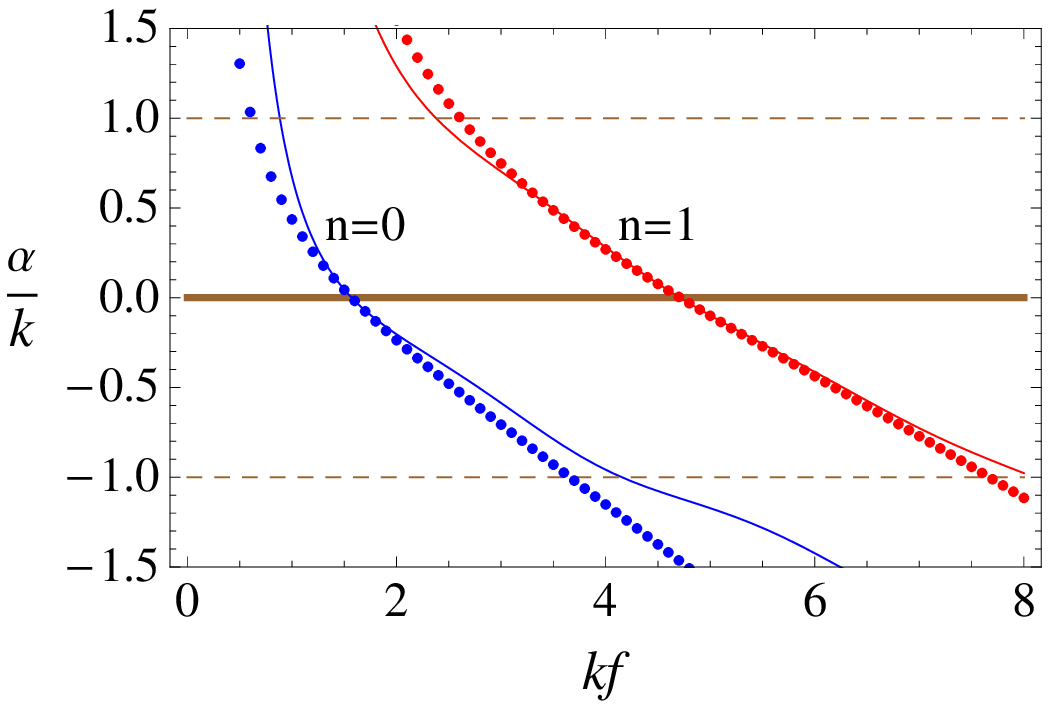}
    \includegraphics[width=0.85\columnwidth]{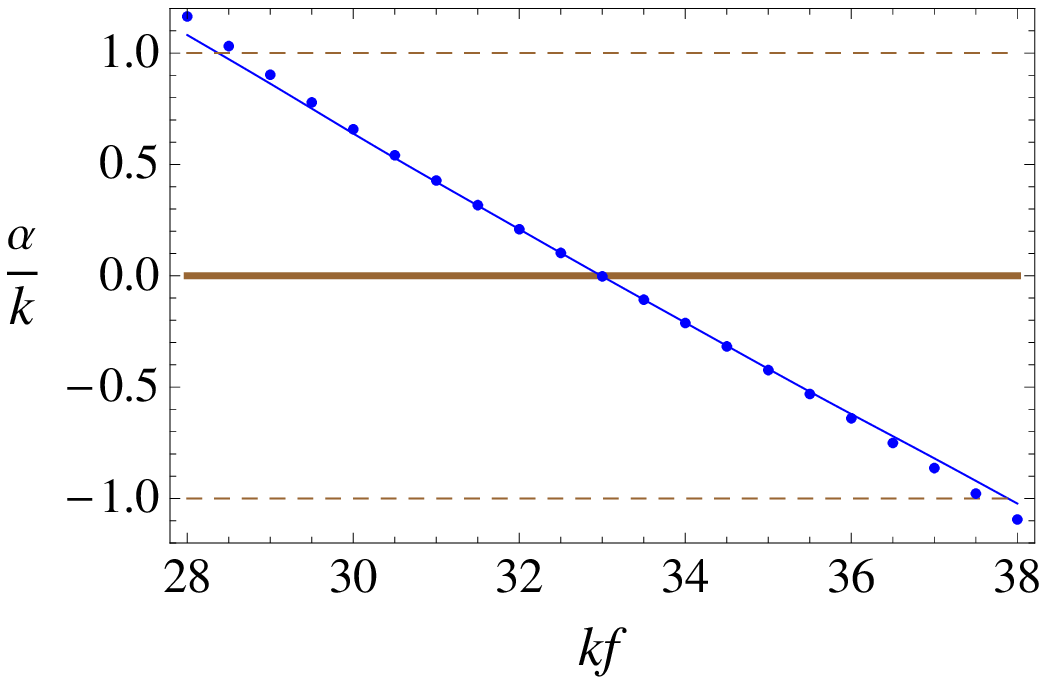}
  \end{center}
  \caption{Quantization of the separation constant $\alpha_n(\omega)$
    as a function of the focal length $f$ of the parabola: exact
    (dots) and analytical (full curves) results for $n=0$ and $n=1$
    (upper figure) and for $n=10$ (lower figure); It is apparent from
    the upper figure that within the range of separation constants
    contributing significantly to the spontaneous decay rate, i.e., 
    $\alpha_n(\omega)/k \in (-1,1)$ (dashed horizontal lines), the
    analytical approximation in Eq.~(\ref{approximation1}) is
    satisfactory even for the lowest possible values of the quantum
    number $n$.}
  \label{Fig:linearization}
\end{figure}

The effective range of integration in Eq.~(\ref{Poisson}) is centered
around $x=0$ with a width $\Delta x = O(1)$. For $x=0$, the mode
function $\chi_{\omega, 0}(\eta)$ and the normalization constant
$\mathcal{N}_{\omega,0}$ are known exactly~\cite{AS}, namely
\begin{eqnarray}
  \chi_{\omega,0} (\eta) & = &  
  \sqrt{\frac{4}{\pi k}}\sin(k \eta/2) \, ,  \nonumber\\
  & &  \\
  \mathcal{N}_{\omega,0} &=& 
  \frac{4}{\pi k} \int_0^{2f}
  d\eta~\frac{\sin^2(k \eta/2)} {\eta} \, .  \nonumber
  \label{solutionexact}
\end{eqnarray}
In the semiclassical limit of large eikonals, i.e., $ k f \gg 1$,
Eq.~(\ref{normal2}) gives the normalization constant
$\mathcal{N}_{\omega,0}$ as
\begin{equation}
  \mathcal{N}_{\omega,0} =  \frac{2\pi}{k}
  \frac{\partial n}{\partial x}(\omega,x=0) \, .
\end{equation}
Therefore, expanding $n(\omega,x)$ around $x=0$, we obtain 
the linear approximation to Eq.~(\ref{solutionexact}), i.e.,
\begin{equation}
  n (\omega,x) = \frac{k f}{\pi} -\frac{1}{2}  +
  x \frac{\partial n}{\partial x}(\omega,x=0)  \, ,
  \label{approximation1}
\end{equation}
where
\begin{equation}
  \frac{\partial n}{\partial x} (\omega, x=0) 
  =  \frac{2\mathcal{S}( k f)}{\pi^2} \, ,
  \label{approximation10}
\end{equation}
and the stability function
\begin{equation}
  \mathcal{S} (u) = \int_0^{u} dy~\frac{\sin^2y}{y} \, ,
  \label{approximation1a}
\end{equation}
which has the asymptotic behavior
\begin{eqnarray} 
 \mathcal{S}(u) &  \xrightarrow [u \gg 1]  & \frac{1}{2}
  \left [ \ln(2u) + \gamma - \frac{\sin 2u}{2u} \right ] +
  O(u^{-2}), \nonumber \\
  & & \\
  \mathcal{S}(u) & \xrightarrow [u \ll 1] & \frac{u^2}{2} + O(u^4) ,
  \nonumber
  \label{approximation1b}
\end{eqnarray}
where $\gamma = 0.5772156649015328606$ is the Euler
constant~\cite{AS}.  The accuracy of this linear approximation can be
appreciated in Fig.~\ref{Fig:linearization}, in which the scaled
separation constant $\alpha/k$ is depicted as a function of the scaled
focal length $kf$. It is apparent that, in range of values of $\alpha$
that contribute significantly to the decay rate in
Eq.~(\ref{Fermiexact}), the linear approximation is quite a
satisfactory description, even in the range of small quantum
numbers $n$.

%%%%%%%%%%%%%%%%%%%%%%%%%%%%%%%%%%%%%%%%%%%%%%%%%%%%%%%%
%%%%%%%%%%%%%%%%%%%%%%%%%%%%%%%%%%%%%%%%%%%%%%%%%%%%%%%

Moreover, by using the integral
\begin{equation}
  \int_{-\infty}^{\infty} dx \frac{x^2}{\sinh^2 x}e^{i\beta x/\pi} =
  \pi^2 \frac{(\beta /2) \coth(\beta/2) -1}{\sinh^2(\beta/2)} \, ,
\end{equation}
we finally obtain in this linear approximation
\begin{eqnarray}
  \frac{\Gamma (\omega_0)}{\Gamma_s(\omega_0)} & = & 
  1  + 2 \sum_{M= 1}^{\infty}3
  \cos[ 2M (u -\pi/2) ] \nonumber \\
  & \times & \left . 
    \frac{2M\mathcal{S}(u) \coth [ 2M \mathcal{S}(u)]-1}
    {\sinh^2 [ 2M \mathcal{S} (u) ]} \right |_{u=k_{0}f} \, .
  \label{linearizationrate0}
\end{eqnarray}
This result allows for a straightforward and elegant interpretation.
The first term on the right-hand side accounts for the free-space
spontaneous decay rate. The terms with $M > 0$ size up the boundary
effects and can be attributed to the repeated reflections of the
emitted photon at the parabolic mirror, with $M$ counting the number
of reflections.  This is noticeable from the characteristic phase
factors in Eq.~(\ref{linearizationrate0}), which appear in integer
multiples of the classical eikonal $2k_{0} f$ characterizing a photon
traveling from the focus, along the symmetry axis, to the mirror and
back again.

With each of those closed orbits an additional phase shift of $\pi$ is
attached, which is equivalent to a Maslov index of
two~\cite{Maslov}. The corresponding amplitude is given by the integer
multiple $M$ of the factor $2\mathcal{S}(k_{0} f)$ that specifies the
stability of the classical trajectories.

%%%%%%%%%%%%%%%%%%%%%%%%%%%%%%%%%%%%%%%%%%%%%%%%%
\begin{figure}[t]
  \begin{center}
    \includegraphics[width=0.85\columnwidth]{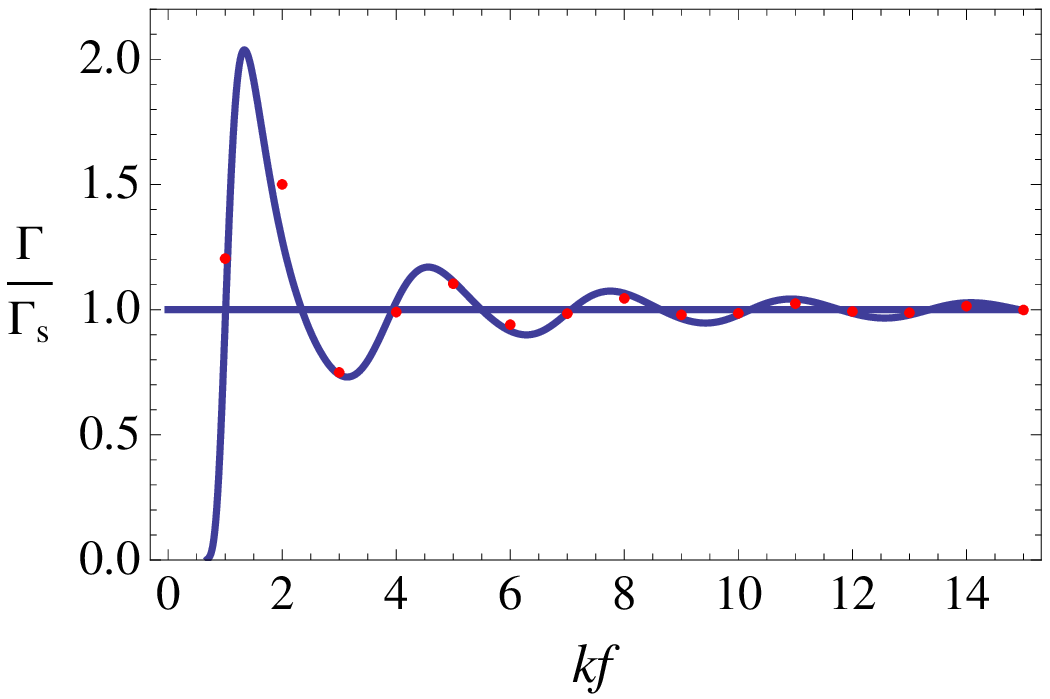}
    \includegraphics[width=0.85\columnwidth]{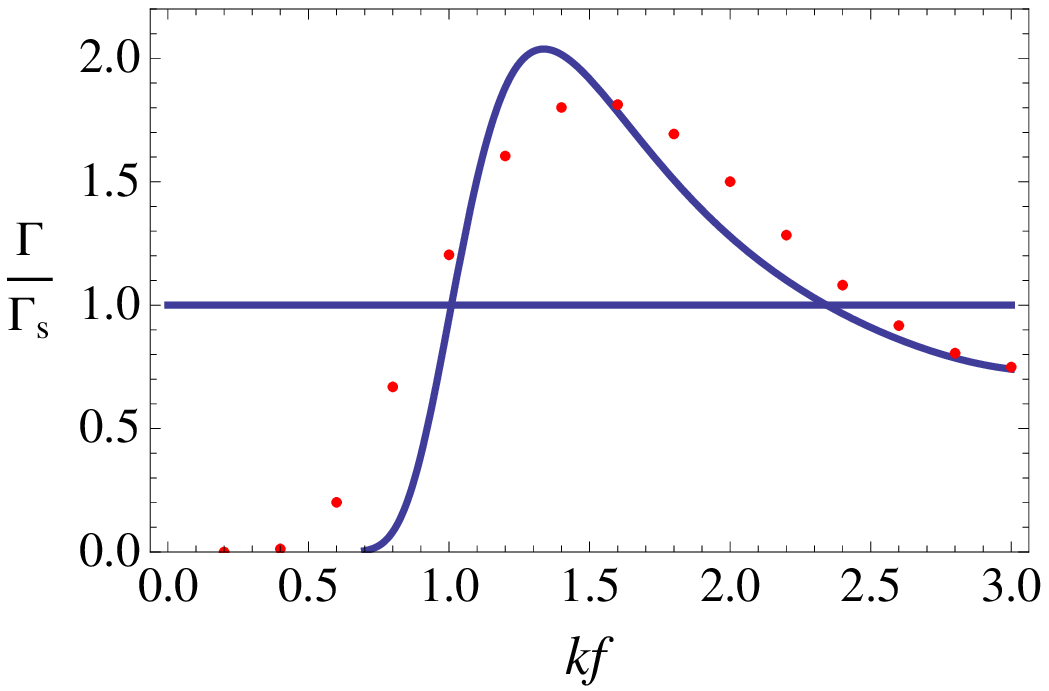}
  \end{center}
  \caption{Scaled spontaneous decay rate
    $\Gamma(\omega_0)/\Gamma_s(\omega_0)$ as a function of the scaled
    focal length $kf$ of the parabola with $k=\omega_0/c$: exact results of
    Eq.~(\ref{Fermiexact}) (dots), semiclassical results (solid) of
    Eq.~(\ref{linearizationrate0}) (we also indicate the
    free-space value as a horizontal line).}
  \label{Fig:Gamma}
\end{figure}
%%%%%%%%%%%%%%%%%%%%%%%%%%%%%%%%%%%%%%%%%%%%%%%%%

The dependence of $\Gamma(\omega_0)$ on the focal length $f$ is shown
in Fig.~\ref{Fig:Gamma}.  It is clear that in the limit of large focal
lengths, i.e. $k_{0} f \gg 1$, $\Gamma(\omega_0)$ eventually tends to
its free-space value $\Gamma_s(\omega_0)$ in an oscillatory
manner. These oscillations evidence the presence of the parabolic
cavity and are satisfactorily described by the $M$-terms in
Eq.~(\ref{linearizationrate0}).

It is also patent from Fig.~\ref{Fig:Gamma} that this approximation
also yields remarkably accurate results even for smaller focal lengths
with $k_{0} f <1$, as long as $n(\omega_0,x=0) \geq 0$.  Furthermore,
we also recognize from Eq.~(\ref{linearizationrate0}) that
$\Gamma(\omega_0)$ vanishes when $k_{0} f \ll 1$.  This reflects that
in those cases the cavity is so small that, at the frequency
$\omega_0$, effectively only the field mode with separation constant
$\alpha_{n=0}(\omega_0)$ is coupled dominantly to the two-level
dipole. From Fig.~\ref{Fig:linearization} one can conclude that
$\alpha_{n=0}(\omega_0)$ tends to infinity for $k_{0} f\ll 1$, so the
factor $[\pi\alpha_n(\omega_0)/ k_0]^2/
[\sinh(\pi\alpha_n(\omega_0)/k_0)]^2$ in Eq.~(\ref{Fermiexact}) tends
exponentially to zero. This situation is in extreme contrast to the
spontaneous photon emission in free space.

\subsection{Dynamics of the spontaneous decay
\label{photondecay}}

We next investigate the dynamics of the spontaneous
photon emission in more detail, to elucidate under which conditions it
can be properly described by an exponential decay.

In the dipole and rotating-wave approximations, the time evolution of
the spontaneous decay is determined by the time-dependent
Schr\"odinger equation with the Hamiltonian (\ref{Hamiltonian}) and
the initial condition (\ref{initialcondition}). Taking advantage of
the explicit form of the semiclassical mode functions of
Sec.~\ref{modefunctions}, the interaction Hamiltonian~(\ref{Hint}) can
be rewritten in the equivalent form
\begin{equation}
  \hat{H}_{AF} =\sum_n \int_0^{\infty}d\omega~ 
  c_{\omega,n}^{\ast} \, 
  | e \rangle \langle g| \, \hat{a}_{\omega,n} + \mathrm{h. \, c.}
\end{equation}
Here, the coupling constants are
\begin{equation}
  c_{\omega,n} = i \sqrt{\frac{2\pi\hbar c}
    {\omega \mathcal{N}_{\omega,n}}} 
  \mathcal{D}(x_n) \, ,
\end{equation} 
with $x_n = c\pi \alpha_n(\omega)/\omega$, and
\begin{equation}
  \mathcal{D}(x) = \sqrt{\frac{3}{\pi^2}
    \frac{x^2}{\sinh^2 x}\frac{\hbar \Gamma_s(\omega)}{2\pi}} \, .
  \label{dipolenew}
\end{equation}
In terms of these quantities, the free-space spontaneous decay rate
can also be recast as
\begin{equation}
  \Gamma_s (\omega_0) =  \frac{2\pi}{\hbar^2}
  \int_0^{\infty}dn~
  \mid c_{\omega_0,n}\mid ^2 = 
  \frac{2\pi}{\hbar}
  \int_{-\infty}^{\infty} dx~| \mathcal{D}(x ) |^2 \, .  
\end{equation}
Consistently with the rotating-wave approximation, we have that
$\Gamma_s (\omega_0)\ll \omega_0$.

For our initial conditions, we can write down
\begin{equation}
  |\psi (t) \rangle = A_e(t) |e\rangle|0\rangle + 
  \sum_n \int_0^{\infty}d\omega~A_{\omega,n}(t) 
  |g\rangle \hat{a}^{\dagger}_{\omega,n}|0\rangle \, , 
  \label{quantumstate1}
\end{equation}
and the resulting Schr\"{o}dinger equation be solved with the help of
Laplace transformation to get
\begin{eqnarray}
  A_{\omega,n}^{(\pm)}(t) &=&  
  \displaystyle
  \pm \frac{1}{2\pi} \int_{-\infty \pm i0}^{\infty \pm i0}
  d\Lambda~e^{-i\Lambda t/\hbar} e^{-iE_gt/\hbar} 
  \frac{c_{\omega,n}}{\Lambda - \hbar\omega}
  \frac{i}{f(\Lambda)} \, , 
  \nonumber \\
  & & 
  \label{Amplitudes}\\
  A_e^{(\pm)}(t) &=& 
  \displaystyle
  \pm \frac{1}{2\pi} \int_{-\infty \pm i0}^{\infty \pm i0}
  d\Lambda~e^{-i\Lambda t/\hbar} e^{-iE_gt/\hbar}
  \frac{i}{f(\Lambda)} \, , 
  \nonumber
\end{eqnarray}
with
\begin{equation}
  f(\Lambda) =    \Lambda - 
  \hbar \omega_0 - \Sigma (\Lambda) \, ,
  \label{denominator}
\end{equation}
and $\Sigma(\Lambda)$ being the self-energy of the two-level system,
viz.
\begin{equation}
  \Sigma (\Lambda) =  \sum_n \int_0^{\infty}d\omega~
  \frac{| c_{\omega,n} |^2}{\Lambda -\hbar \omega} \, .
  \label{selfenergy}
\end{equation}
The $\pm$ signs refer to the retarded ($+$) and advanced ($-$)
solutions valid for ${\rm sgn}(t) = \pm 1$. The notation $\pm i0$
indicates that the integration has to be performed in the complex
$\Lambda$ plane parallel to the real axis with an infinitesimal
positive ($+$) or negative ($-$) imaginary offset.

\subsubsection{The self-energy in the semiclassical approximation}

To perform the $\Lambda$-integrations involved in
Eqs.~(\ref{Amplitudes}), we have first to determine the
$\Lambda$-dependence of the self-energy function $f(\Lambda)$ in the
region around $\Lambda \approx \hbar\omega_0$.  To achieve this, we
redraft Eq.~(\ref{selfenergy}) with the help of the Poisson summation
formula as
\begin{equation}
  \Sigma(\Lambda)=  - \sum_{M=-\infty}^{\infty} \int_{-\infty }^{\infty}dx
  \int_0^{\infty} d\omega~\frac{\partial n(\omega,x)}{\partial x}
  \frac{| c_{\omega n} |^2}{\hbar \omega - \Lambda}
  e^{i2\pi M n(\omega,x)} \, . 
  \label{SigmaPoisson1}
\end{equation}
Thereby, we have used the smooth real-valued function $n(\omega ,x)$
given by Eq.~(\ref{approximation1}).  The imaginary part of
$n(\omega,x)$ tends to plus (minus) infinity for large $\omega$ with
positive (negative) imaginary part, whence we obtain the semiclassical
approximation
\begin{eqnarray}
  \Sigma^{\pm}(\Lambda) & =&
  \Sigma(\Lambda \pm i0) =  \hbar \delta \omega
  \mp i\frac{\hbar \Gamma_s(\Lambda/\hbar)}{2}  
  \nonumber  \\
  & \times & \left [
    1 +  \frac{6}{\pi^2} \sum_{M=1}^{\infty}
     \int_{-\infty}^{\infty} dx~\frac{x^2}{\sinh^2 x}
    e^{\pm i2\pi M n(\Lambda/\hbar, x)}
  \right ] \, , 
  \label{selfenergypath}
\end{eqnarray}
for real values of $\Lambda$. Herein, $\delta \omega$ is the resonant
contribution of the Lamb shift of the two-level transition
$|g\rangle~\to~|e\rangle$ originating from the real part of the
$M=0$ contribution in Eq.~(\ref{SigmaPoisson1}). Henceforth, we assume
that this contribution is incorporated in a renormalized transition
frequency $\omega_0$ in which the complete Lamb shift is taken into
account in second-order perturbation
theory~\cite{LambShift}.

The contribution of a particular value of $M\geq 1$ in
Eq.~(\ref{selfenergypath}) can be attributed to $M$ reflections of a
photon path at the parabolic mirror; each photon path being associated
with a particular value of the separation constant $x \in
(-\infty,\infty)$.

By adopting the approximation~(\ref{selfenergypath}), the
$\Lambda$-integration involved in Eqs.~(\ref{Amplitudes}) can be
performed in two complementary ways. In the dressed-state
representation, this integration is evaluated using residue
calculus. Alternatively, this can be directly performed with the help
of a photon path representation of the integrand. In what follows, we
scrutinize both options.

\subsubsection{The dressed-state representation in 
the pole approximation
\label{poleapproximation}}

The poles of the integrands in Eq.~(\ref{Amplitudes}) [which stem from
the zeros of $f(\Lambda)$] yield the complex-valued dressed energies
of the strongly-coupled atom-field system.  The time dependent quantum
state~(\ref{quantumstate1}) can thus be expressed as a sum of
contributions of all these dressed states.

In all these $\Lambda$-integrations the dominant contributions are
expected to arise from values of $\Lambda \approx \hbar
\omega_0$. According to Eq.~(\ref{selfenergypath}), the characteristic
values of the self-energy $\Sigma^{\pm}(\Lambda)$ are of the order of
$O(\hbar \Gamma_s(\Lambda/\hbar))$.  Therefore, as long as the
self-energy is a slowly-varying function of $\Lambda$ around $\Lambda
\approx \hbar \omega_0$, i.e., whenever
\begin{equation}
  \hbar \Gamma_s(\omega_0)~ \left | 
    \frac{\partial \Sigma^{\pm}   (\Lambda)}
    {\partial \Lambda} \right |_{\Lambda = \hbar \omega_{0}} \ll 1 \, ,
  \label{slowapprox}
\end{equation}
we can approximate $\Sigma^{\pm}(\Lambda)$ by its value at
$\Lambda = \hbar\omega_0$.  In this case, the equation $f(\Lambda) = 0$
has only one solution, namely
\begin{equation}
  \Lambda_0 = \hbar \omega_0 + \Sigma^{\pm}(\hbar \omega_0) \, .
\end{equation}
In this pole approximation~\cite{poleapproximation}, the dressed-state
representation~(\ref{Amplitudes}) reads
\begin{equation}
  A^{(\pm)}_e(t) = e^{-i [E_g + \hbar \omega_0 +\Delta  (\omega_0)]t/\hbar} 
  e^{- | t | \Gamma(\omega_0)/2} \, , 
  \label{poleapprox}
\end{equation}
for both the retarded ($t\in[0,\infty)$) and advanced ($t\in
(-\infty,0]$) dynamics of the state $|e\rangle$.  The quantity $\Delta
(\omega_0) = {\rm Re} [ \Sigma^{\pm}(\hbar \omega_0) ]$ represents the
resonant energy shift induced by the spontaneous emission.
Equation~(\ref{poleapprox}) hints at an exponential decay of the
probability amplitude $A_e(t)$, with a rate $\Gamma(\omega_0)$.  This
is a consequence of the rotating-wave approximation, which involves an
averaging over times scales of the order of $1/\omega_0$.  At very
short times (say, of the order of $1/\omega_0$ or less), this pole
approximation breaks down and deviations from an exponential decay may
occur~\cite{expodecaydeviation}.

From the linear estimate (\ref{approximation1}) we conclude that the
semiclassical approximation (\ref{selfenergypath}) is valid as long as
$k_{0} f \gg 1$, so that many modes are excited by the spontaneous
emission. In addition, $f$ must also be small enough so that
inequality (\ref{slowapprox}) is fulfilled. This latter condition
implies $f \ll c/\Gamma_s(\omega_0)$ and states that the focal length
still has to be significantly smaller than the typical length $\Delta
l = c/\Gamma_s(\omega_0)$ of the spontaneously generated photonic wave
packet in free space. Physically speaking, this condition implies that
the repeated reflections of the photon wave packet at the parabola
overlap significantly in space-time, so that they cannot be resolved
and thus interfere.  This interference gives rise to the oscillations
of the spontaneous decay rate $\Gamma(\omega_0)$, which have already
been investigated in subsection~\ref{Rate}.

\subsubsection{The semiclassical photon-path representation}

As soon as the smoothness condition (\ref{slowapprox}) is no longer
fulfilled, the $\Lambda$-integrations involved in
Eqs.~(\ref{Amplitudes}) have to be evaluated by more sophisticated
means. According to our previous
considerations, this happens if the focal length is so large that $f >
c/\Gamma_s(\omega_0)$. In this instance, the contributions from the
repeated reflections of the photon at the parabola are separated
sufficiently well in space-time, so that they can be distinguished by
appropriate measurements.

In this regime, a systematic photon-path representation of the
amplitude $A^{(\pm)}_e(t)$ is convenient not only from the
computational, but also from the physical point of view.  In
the spirit of the path-integral approach, $A^{(\pm)}_e(t)$ can
be appropriately represented as a sum of amplitudes associated with
all photon paths which start and end at the position of the two-level
system and which are reflected repeatedly at the cavity.

To obtain that representation, we start from the semiclassical
approximation of the self-energy given in
Eq.~(\ref{selfenergypath}). After some direct computations, one gets
\begin{widetext}
  \begin{eqnarray}
    \frac{1}{f(\Lambda \pm i0)} &=& 
    \frac{1}{f_0(\Lambda \pm i0)} \pm 
    i2\pi
    \frac{1}{f_0(\Lambda \pm i0)}
    \int_{-\infty}^{\infty} dx_1~
    \int_{-\infty}^{\infty} dx_2~
    \mathcal{D}(x_1)\mathcal{Y}^{\pm}(x_1,x_2) 
    e^{\pm i 2\pi n(\Lambda/\hbar, x_2)}\mathcal{D}(x_2)
    \frac{1}{f_0(\Lambda \pm i0)}
    \label{finverse}
  \end{eqnarray}
\end{widetext}
where
\begin{eqnarray}
  f_0(\Lambda \pm i0) &=& 
  \Lambda - \hbar \omega_0 \pm i\frac{\hbar \Gamma_s(\Lambda/\hbar)}{2},
  \nonumber\\
  & & \\
  \mathcal{Y}^{\pm}(x_1,x_2) &=&  - \delta(x_1 - x_2) \nonumber \\ 
  & + & \int_{-\infty}^{\infty} dx_3~\mathcal{Y}^{\pm}(x_1,x_3) 
  e^{\pm i 2\pi n(\Lambda/\hbar,x_3)}
  {S}^{\pm}(x_3,x_2)  \, ,  \nonumber
  \label{photonpathrepresentation}
\end{eqnarray}
and the scattering $S$-matrix given by
\begin{equation}
  {S}^{\pm}(x_3,x_2) =\delta (x_3 - x_2) 
  \mp 2 i \pi \mathcal{D}(x_3)\frac{1}{f_0(\Lambda \pm i0)}
  \mathcal{D}(x_2)  \, .
\end{equation} 

Let us set out the (generalized) basis vectors $\{| x \rangle\}$ in
the Hilbert space of square-integrable functions of the separation
constant $x \in \mathbb{R}$.  Using Eq.~(\ref{dipolenew}) and defining
the dipole vector $|{\cal D}\rangle$ by
\begin{equation}
  |\mathcal{D}\rangle = \int_{-\infty}^{\infty} dx~\mathcal{D}(x) |x\rangle, 
  \qquad
  \langle \mathcal{D}| :=  \int_{-\infty}^{\infty} dx~\mathcal{D}(x) \langle x|,
\end{equation}
Eq.~(\ref{finverse}) can be conveniently recast as
\begin{eqnarray}
  \label{help10}
  \frac{1}{f(\Lambda \pm i0)} &=& \frac{1}{f_0(\Lambda \pm i0)} \mp
  i 2\pi\frac{1}{f_0(\Lambda \pm i0)} \nonumber \\
  & \times & \sum_{M=0}^{\infty}
  \langle \mathcal{D}|
  \left(e^{\pm i 2\pi \mathbf{n}(\Lambda/\hbar)} 
    \mathbf{{S}}^{\pm}\right)^M 
  e^{\pm i 2\pi \mathbf{n}(\Lambda/\hbar)}  |\mathcal{D}\rangle
  \nonumber \\
  & \times &
  \frac{1}{f_0(\Lambda\pm i0)} \, ,
\end{eqnarray}
with the scattering operator
\begin{equation}
  \mathbf{{S}}^{\pm} =  \openone \mp 2 i \pi |{\cal
    D}\rangle\frac{1}{f_0(\Lambda\pm i0)}\langle \mathcal{D}| \, .
  \label{scattering-T-matrix}
\end{equation}
The quantity
\begin{equation}
  e^{\pm i 2\pi \mathbf{n}(\Lambda/\hbar)}
  = \int_{-\infty}^{\infty} dx~|x\rangle \langle x| 
  e^{\pm i 2\pi n(\Lambda/\hbar,x)}
\end{equation}
encodes the phase accumulated by a photon during all closed paths
starting and ending at the focus.

For sufficiently large focal lengths $f$, for which the pole
approximation is not applicable but for which the linear
approximation still applies,  $A^{(\pm)}_e(t)$ can be evaluated term by term with the
help of Eq.~(\ref{help10}). The contributions with $M\leq 2$, for
example, are explicitly given by
\begin{widetext}
  \begin{eqnarray}
    \label{photonpath10}
    A^{(\pm)}_e(t) &=&
    e^{-i(E_g/\hbar + \omega_0 )t -\mid t\mid \Gamma_s(\omega_0)/2}
    + \sum_{M=1}^{2}  \Theta(\mid t\mid - MT)~
    e^{-i(E_g/\hbar + \omega_0) t}~e^{\pm i2\pi M n(\omega_0,x=0)} 
 \nonumber \\
    & \times &
    \sum_{k=0}^{M-1} {M-1\choose  k}
    \frac{(-1)^{k+1}}{(k+1)!}
    \left[(\mid t\mid  - MT) \Gamma_s(\omega_0) \right]^{k+1}
    e^{ - (\mid t\mid - MT)\Gamma_s(\omega_0)/2} \nonumber\\
    & \times &\left(
      \int_{-\infty}^{\infty} dx~\frac{3x^2}{\pi^2\sinh^2 x}
      e^{\pm ix2\pi (M -k) (\partial n/\partial x)(\omega_0,x=0)}
    \right)
    \left ( 
     \int_{-\infty}^{\infty} dx~\frac{3x^2}{\pi^2\sinh^2x}
      e^{\pm ix2\pi (\partial n/\partial x)(\omega_0,x=0)}
    \right )^k + \cdots 
  \end{eqnarray}
\end{widetext}
In this expression 
\begin{equation}
T := 2\pi \hbar \frac{\partial n (\omega_0,x=0)}{\partial \Lambda} =
\frac{2f}{c}
\end{equation}
is the classical period of the closed photon path with separation
constant $x=0$, which starts at the focus of the parabola and extends
along the symmetry axis to the vertex and back again.

The various terms appearing in Eq.~(\ref{photonpath10}) allow again
for a sensible interpretation. In the retarded solution $(+)$, for
example, the very first term on the right-hand side expresses the
spontaneous decay process of the excited state $|e\rangle$ in the
absence of the cavity and is governed by free-space decay rate
$\Gamma_s(\omega_0)$.  The remaining terms with $M\geq 1$ describe the
time dependence of the probability amplitudes for the photon after $M$
reflections. The $M$th contribution represents a process in which a
photon is emitted at time $t=0$ and is reabsorbed again at a time $t
\geq MT$. This photon accumulates a phase of $2\pi M n(\omega_0,x=0)$,
which is the eikonal associated with the classical path with
separation constant $x=0$.  Furthermore, the photon can be scattered
$0\leq k \leq (M-1)$ times during its intermediate returns to the
focus of the parabolic cavity.

The stability of the path with $x=0$ is described by the
characteristic quantity $\partial n/\partial x (\omega_0,x=0)$
[compare with Eq.~(\ref{approximation1a})].  If this path were
stable, i.e., $\partial n/\partial x (\omega_0,x=0) = 0$, the
representation~(\ref{photonpath10}) would reduce to
previous results for a spherically symmetric
cavity, in which all relevant photon paths are stable and their
contributions add up in phase~\cite{Alber:1992qr}.

As far as the time evolution of $A_e^{(\pm)}(t)$ is concerned, two
different regimes may be distinguished. If the classical period is
significantly smaller than the free-space decay, i.e., $T = 2f/c \ll
1/\Gamma_s(\omega_0)$, the probability amplitudes associated with
different {photon bounces $M\geq 1$ overlap in time significantly and
  cannot be distinguished in the evolution. In this case,
  $A_e^{(\pm)}(t)$ leads to an exponential decay with the modified
  rate $\Gamma(\omega_0)$, which has been discussed in detail in
  Eqs.~(\ref{Fermiexact}) and (\ref{linearizationrate0}).  In the
  opposite limit of long classical periods, i.e., $T = 2f/c \gg
  1/\Gamma_s(\omega_0)$, the contributions associated with different
  returns $M\geq 1$ are well separated in time and the overall time
  evolution of $A_e^{(\pm)}(t)$ is modified significantly.

%%%%%%%%%%%%%%%%%%%%%%%%%%%%%%%%%%%%%%%%
  \begin{figure}[t]
    \includegraphics[width=0.85\columnwidth]{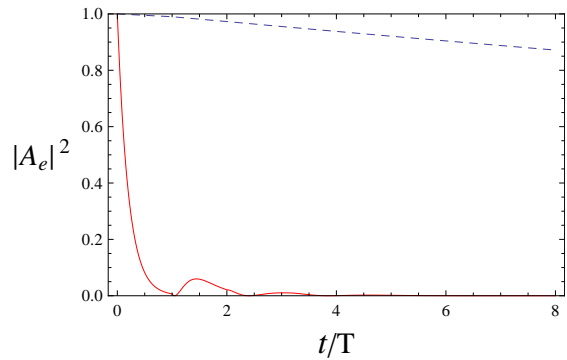}
    \caption{Time evolution of the probability $\mid
      A_e^{(\pm)}(t)\mid^2$. The interaction time $t$ is plotted
      in units of the classical period $T$ of the photon path
      connecting the two-level system at the focus of the parabola
      with its vertex. The parameters are $n = k_{0} f/\pi -
      1/2 = 0$, $\Gamma_s(\omega_0) T = 0.01$ (dashed curve) and
      $\Gamma_s(\omega_0) T = 5$ (full curve).}
    \label{Fig:Timeevolution}
  \end{figure}
%%%%%%%%%%%%%%%%%%%%%%%%%%%%%%%%%%%%%%%

  The time evolution of the probability $| A_e^{(\pm)}(t) |^2$}
is depicted in Fig.~\ref{Fig:Timeevolution}.  For typical free-space
decay rates of the order of $10^9$~s$^{-1}$, focal lengths
significantly larger than $10$~cm would be required. Furthermore, in
view of the stability properties of the photon path with $x=0$,
already for small values of the quantization number $n$ the
contributions of repeated returns of the photon with $M \geq 1$ are
suppressed significantly. So, in view of current technological
capabilities, the experimental observation of the modified spontaneous
decay as described by Eq.~(\ref{photonpath10}) is challenging.

To sum up this discussion, with the present values in our experimental
setup (focal length $f=2.1$ mm and wavelength of $369$ nm), one can reach
the typical condition for strong coupling. It follows then that there should
be several phenomena which can be observed in this limit, as reviewed in Ref. \cite{LeSo}

\section{Photon dynamics 
\label{OnePhoton}}

Although the mean values of the electric and magnetic field strengths
vanish for any one-photon state, their fluctuations may be seen as
stemming from an effective one-photon amplitude. In the long-time
limit, this latter amplitude consists of an asymptotic wave
propagating along the symmetry axis of the parabola with a
characteristic transversal spatial modulation and polarization pattern
and of a spherically outgoing wave whose amplitude vanishes far from
the focus of the parabola.  The multiple reflections of the
spontaneously-emitted photon manifest themselves in the transversal
spatial modulations.

An especially interesting quantity to ascertain the dynamics of this
single-photon state is the normally-ordered electric field correlation
function. Taking into account the time evolution given by
Eq.~(\ref{quantumstate1}), we can write \cite{StoAlLe}
\begin{eqnarray}
  & \langle \psi (t) |: \hat{E}_{k} (\mathbf{x})~\hat{E}_{\ell} (\mathbf{x}) :
  |\psi(t)\rangle = & 
  \left (
    \mathbf{e}_{k}\cdot \nabla \times \mathbf{F}^{\pm}(\mathbf{x}, t)\right) ~
  \nonumber \\
  & \left(
    \mathbf{e}_{\ell} \cdot \nabla  \times \mathbf{F}^{\pm*}(\mathbf{ x}^{\prime}, t) 
  \right )
  + \mathrm{c. \, c.} \, , &
  \label{correlation}
\end{eqnarray}
where $\mathbf{e}_{k, \ell}$ are Cartesian unit vectors and the
effective one-photon amplitude reads
\begin{equation}
  \mathbf{F}^{\pm} (\mathbf{x}, t)  = -i \sum_{n}
  \int_0^{\infty}d\omega~\sqrt{\frac{\hbar \omega}{2\epsilon_0}}
  A^{\pm*}_{\omega.n}(t) \mathbf{G}^{\ast}_{\omega,n}(\mathbf{x}).
  \label{onephotonamplitude}
\end{equation}
According to Glauber's theory~\cite{Glauber}, this (and analogous
normally-ordered higher-order correlation functions) can be measured
by optical photodetection.

In general, the one-photon amplitude (\ref{onephotonamplitude}) has to
be evaluated numerically.  However, we can grasp its basic space-time
dependence, if we concentrate on the long-time limit $\Gamma(\omega_0)
|t| \gg 1$, for which
\begin{equation}
  A^{(\pm)}_{\omega,n}(t) = e^{-i\omega t}e^{-iE_gt/\hbar}
  \frac{c_{\omega,n}}{\hbar(\omega - \omega_0) -
    \Sigma^{\pm}(\hbar \omega)}\, ,
\end{equation}
where we have used Eq.~(\ref{Amplitudes}) and neglected exponentially
small terms.  Provided the semiclassical relation (\ref{normal2}) is
applicable, the one-photon amplitude~(\ref{onephotonamplitude}) can be
rewritten, using the Poisson summation formula, as a sum of all
possible photon paths originating at the focus, i.e.,
\begin{eqnarray} 
\mathbf{F}^{\pm}(\mathbf{x},t) &=& -
  \mathbf{e}_{\varphi} \sum_{M= - \infty}^{\infty}
  \int_0^{\infty}d\omega~\int_{-\infty}^{\infty}dx~
  \frac{x}{\sinh x} \nonumber \\
  & \times & \sqrt{\frac{3\Gamma_s(\omega)\hbar^3\omega} {16\epsilon_0
      c\pi^5}} \, \frac{\chi_{\omega.n}(\xi)}{\sqrt{\xi}}
  \frac{\chi_{\omega,n}(\eta)}{\sqrt{\eta}}
  \nonumber \\
  & \times & e^{i(\omega + E_g/\hbar)t} \frac{e^{i2\pi M
      n(\omega,x)}}{\hbar(\omega - \omega_0) - \Sigma^{\mp}(\hbar
    \omega)}.
  \label{onephotonamplitude1}
\end{eqnarray}
The dominant contribution to the integral over $x$ comes from a small
neighbourhood around $x=0$, so we can use the linear approximation
(\ref{approximation1}) for $n(\omega,x)$. Besides, in the radiation
zone, i.e., far from the focus of the parabola, the mode functions
$\chi_{\omega,n}(\xi)$ and $\chi_{\omega,n}(\eta)$ can be replaced by
their asymptotic expressions (see the Appendix).  Finally, if the focal
length $f$ is not too large, i.e.,  $2f\Gamma(\omega_0)/c \ll 1$, the
$\omega$-integration can be performed with the pole
approximation. With all this in mind, the final result turns out to be
\begin{widetext}
  \begin{eqnarray}
    \mathbf{F}^{\pm}(\mathbf{x},t) &=& \pm i \frac{\mathbf{e}_{\varphi}}{\sqrt{\xi \eta}}
    \sqrt{\frac{3\Gamma_s(\omega_0)\hbar c}{4\epsilon_0 \pi^5
        \omega_0}} e^{i E_g t/\hbar} \sum_{M\geq 0}
    e^{\mp i2\pi M n(\omega_0,x=0)}  \nonumber\\
    &\times & \left \{ \Theta \left ( | t | - \frac{\xi + \eta}{2c} -
        M T \right ) \frac{\pi^2 e^{\pm i\omega_0 [ | t | -
          (\xi + \eta)/(2c)]} \; e^{-\Gamma(\omega_0) [ | t | - (\xi +
          \eta)/(2c)]/2}}
      {2\cosh^2[-\ln\sqrt{\xi/\eta} + 2 M \mathcal{S}(k_{0} f)]}  \right. \nonumber\\
    & - &\left.  \Theta \left ( | t | - \frac{\xi - \eta}{2c} - M
        T \right ) \frac{\pi^2 e^{\pm i \omega_0 [| t | -
          (\xi - \eta)/(2c)]} \, e^{-\Gamma(\omega_0)[ | t | - (\xi -
          \eta)/(2c)]/2}} {2\cosh^2[-\ln\sqrt{\eta\xi/(2f)^2} + 2
        (M-1) \mathcal{S}(k_{0} f)]} \right\} \, .
    \label{onephotonamplitude2}
  \end{eqnarray}
\end{widetext}
In the derivation of Eq.~(\ref{onephotonamplitude2}), we have used the
relation
\begin{equation}
  \int_{-\infty}^{\infty}dx~\frac{x}{\sinh x} e^{i\beta x/\pi} =
  \frac{\pi^2}{2\cosh^2(\beta/2)} \, ,
\end{equation}
and we have neglected exponentially small terms. In the limit
$\Gamma(\omega_0) 2f/c \ll 1$, which we are considering here, the unit step
functions $\Theta (u)$ associated with different $M$-values have
almost identical support.  For $\mathcal{S}(\omega_0 f/c) \gg 1$, in
Eq.~(\ref{onephotonamplitude2}) only contributions with $M=0$ for the
first term in curly brackets and with $M=1$ for the second term in
curly brackets are significant so that the one-photon amplitude
simplifies to
  \begin{eqnarray}
    F^{\pm}(\mathbf{x}, t) &=& 
    \pm i \mathbf{e}_{\varphi}
    \sqrt{\frac{3\Gamma_s(\omega_0)\hbar c}{16\epsilon_0 \pi \omega_0}}
    e^{i E_g t/\hbar} \nonumber\\
    & \times &
    \left (
      \Theta (\mid t\mid - r/c)
      e^{\pm i\omega_0(\mid t\mid - r/c)} e^{-\Gamma(\omega_0)(\mid t\mid - r/c)/2}
      ~\frac{\rho}{r^2} \right . \nonumber\\
    &- &
      \Theta \left ( |t | - z/c - T\right ) 
      e^{\pm i\omega_0(\mid t\mid - z/c)} e^{-\Gamma(\omega_0)(\mid
        t\mid - z/c)/2} \nonumber \\ 
& \times & \left . e^{\mp i2\pi n(\omega_0,x=0)}~
      \frac{2}{f}\frac{\rho/(2f)}{\left \{  1 + [ \rho/(2f) ]^2 \right \}^2}
    \right ) \, .
  \end{eqnarray}
This expression could also be obtained by evaluating the one-photon
amplitude directly with the help of multidimensional
Jeffreys-Wentzel-Kramers-Brillouin (JWKB) methods~\cite{Maslov}. This indicates that the residual terms
in Eq.~(\ref{onephotonamplitude2}) can be attributed to diffraction
phenomena and to photon scattering by the atom. In particular, these
contributions become important for values of the stability function
${\cal S}(k_0 f )$ of the order of unity.

%%%%%%%%%%%%%%%%%%%%%%%%%%%%
\begin{figure}[b]
  \includegraphics[width=0.85\columnwidth]{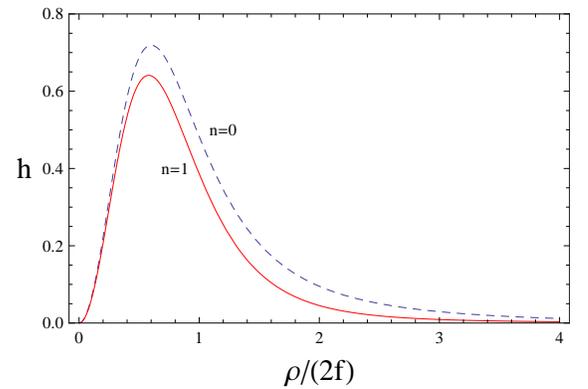}
  \caption{Asymptotic transverse energy distribution $h(\rho/(2f)) =
    \pi (2f)^2 I(y){\Gamma(\omega_0)}/{\Gamma_s(\omega_0)}$ of the
    one-photon quantum state for two different values of the
      quantization constant $n=0$ (dashed line) and $n=1$
    (full line).}
  \label{Fig:photondistribution}
\end{figure}
%%%%%%%%%%%%%%%%%%%%%%%%%%%%%%%%%%%%%%

Starting from Eq.~(\ref{onephotonamplitude2}), it is straightforward
to demonstrate that in an asymptotic plane $z \to \infty$, the photon
transversal energy density is
\begin{eqnarray}
  \int_{-f}^{\infty}dz~
  \frac{\epsilon_0}{2} \langle \psi(t)| : 
  \hat{\mathbf{E}}^2(\mathbf{x}) + c^2\hat{\mathbf{B}}^2(\mathbf{x}):|\psi(t)\rangle 
  =  \hbar \omega_0 I(y) \, ,
  \nonumber \\
\end{eqnarray}
with $y= [ \rho/(2f)]^2$ and with the planar energy distribution
\begin{eqnarray}
  I(y) &=& 
  \frac{\Gamma_s(\omega_0)}{\Gamma(\omega_0)}
  \frac{1}{4 \pi f^2}\left \{
    6 \frac{y}{(1+y)^4} \right . \nonumber \\
  & + & \left . 12 \sum_{M=1}^{\infty} \cos[ 2M(k_{0} f - \pi/2)]
    \frac{ye^{2M v}}{(1+y)^2(e^{2Mv} + y)^2} \right \} \, , 
  \nonumber \\
  \label{transversedistribution}
\end{eqnarray}
with $v = 2\mathcal{S}(k_{0} f)$.  With the help of
Eq.~(\ref{linearizationrate0}) one can easily check that
\begin{equation}
  \int_0^{2\pi}d\varphi \int_0^{\infty}d\rho~\rho I(y) =1 \, ,
\end{equation}
consistent with the fact that, within the rotating-wave approximation,
the total energy of the one-photon state is $\hbar \omega_0$.

According to Eq.~(\ref{transversedistribution}), the terms with
  $M \geq 1$ contribute the most when the quantization condition
  $k_{0} f = \pi (n + 1/2)$ is fulfilled for $n \in {\mathbb N}$.  In
Fig.~\ref{Fig:photondistribution} the transversal energy distribution
is depicted for the frequencies corresponding to the two lowest
  integer quantization values $n= 0$ and $n=1$.  For
$n=0$, the contributions of repeated reflections at the parabolic
boundary slightly modify the transversal energy distribution.
However, already for $n=1$,
it turns out that, due to the instability of the photon path
determined by the parameter $v$, the contributions of the terms with
$M \geq 1$ are negligible in the transverse energy distribution.

We also point out that we have recently demonstrated the experimental
generation of temporal modes allowing for an efficient coupling of single 
photons and single two-level systems in free space using a deep parabolic mirror \cite{34},
which confirms the theory developed here.

\section{Concluding remarks}

We have explored the dynamics of the spontaneous emission by a
two-level atom at the focus of a parabolic cavity. Concentrating on
the optical regime, we have determined the time evolution of both the
atomic system and the photon in the dipole and rotating-wave
approximations.  We have investigated also the advanced solution, for
it approaches, at a particular time ($t=0)$, a quantum state in which
the atom is in its excited state and the field in its ground (vacuum)
state. Thus, it describes a physical situation in which the atom is
excited with near certainty by a single photon in a multimode
scenario.

By taking into account the vectorial character of the electromagnetic
field, we have demonstrated that the photon exchange can be described
in a physically transparent way with the help of semiclassical
methods. Thereby, the observables of interest are represented as sums
of probability amplitudes associated with repeated reflections of the
photon at the boundary of the parabolic cavity and with its repeated
resonant scatterings by the atom.  This semiclassical description does
not only yield a quantitatively adequate description of this physical
exchange in the limit of short wavelengths, but also constitutes a
satisfactory approximation in the opposite limit of long wavelengths.

Bearing in mind the current experimental activities aiming at the
realization of quantum repeaters, half-open cavities, such as the one
discussed in this paper, offer interesting perspectives for coupling
an elementary material qubit almost perfectly to the electromagnetic
radiation field even in extreme multimode scenarios. Experimental work
in that direction is in progress in our laboratory.

\begin{acknowledgments}
  Financial support from the EU FP7 (Grant No. Q-ESSENCE), the BMBF
  (project QK{\_}QuOReP), the Spanish DGI (Grant No. FIS2011-26786), and
  the UCM-BSCH program (Grant No. GR-920992) is acknowledged. It is also a pleasure to thank A. R. P. Rau and
  M. Sondermann for stimulating discussions. MS is supported by the EU 7FP Marie Curie Career Integration 
  Grant No. 322150 “QCAT”, NCN grant No. 2012/04/M/ST2/00789, FNP Homing Plus project No. HOMING PLUS/2012-5/12 
  and MNiSW co-ﬁnanced international project No. 2586/7.PR/2012/2.
\end{acknowledgments}

\appendix
\section{}
In this appendix, details concerning the separation of the vectorial
Helmholtz equation in parabolic coordinates and the frequency
normalization of the mode functions are summarized.

The parabolic coordinates $(\xi, \eta, \phi)$ are defined through
\begin{equation}
  \label{eq:parabolic}
  x =  \sqrt{\xi \eta} \cos \phi \, , 
  \quad 
  y =  \sqrt{\xi \eta} \sin \phi \, ,
  \quad
  z = \frac{1}{2} (\xi - \eta) \, ,
\end{equation}
where $0 \le \xi$, $\eta < \infty$, and $0\le \phi <2\pi$. The surfaces
$\eta = {\rm constant} $ are paraboloids of revolution about the
positive $Z$ axis, having their focal point at the origin, while the
surfaces $\xi = $ const are directed along the negative $Z$ axis. The
plane $z=0$ corresponds to the condition 􏰸$\xi = \eta$.

In these coordinates, the Laplacian operator governing the vectorial
Helmholtz equation (\ref{Helmholtz}) is given by
\begin{equation}
  \Delta =  \frac{4}{\xi +\eta} 
  \left(
    \frac{\partial }{\partial \xi}
    \xi \frac{\partial }{\partial \xi} +
    \frac{\partial }{\partial \eta}
    \eta \frac{\partial }{\partial \eta} 
  \right ) + 
 \frac{1}{\xi \eta} \frac{\partial^2}{\partial \phi^2} \, .
\end{equation}
This implies that mode functions of the form of Eq.~(\ref{modeansatz})
fulfill Eqs.~(\ref{Coulomb}). According to Eq.~(\ref{separation})
these mode functions involve Coulomb functions of zero angular
momentum $L$, $F_{L=0} ( \mu , \rho )$. We have the asymptotic 
limit (for large $\rho$) 
\begin{equation}
  F_{L=0} ( \mu , \rho ) \sim  \sin \Phi (\mu, \rho ) \, ,
\end{equation}
with the Coulomb phase
\begin{equation}
  \Phi(\mu, \rho ) = \rho - \mu \ln (2 \rho) + \arg  \Gamma  ( 1 + i
  \mu )  \, 
  \label{Coulombphase}
\end{equation}
For $\mid \mu \mid \ll 1$, the corresponding
asymptotic expression reads
\begin{equation}
  F_{L=0} (\mu, \rho )  \sim 
  \rho   \sqrt{\frac{e^{ \pi \mu} \pi \mu}
   {\sinh (\pi \mu)}} \, .
\end{equation}
Note that in our case, $\rho = \omega \eta/(2c)$ and
$\mu = - \alpha c/\omega$.

If $\mu$ is of the order of unity or less, then, consistent with the
linear approximation~(\ref{approximation1}), the argument of the $\Gamma$
function appearing in the Coulomb phase~(\ref{Coulombphase}) can be
approximated by $\arg \Gamma (1 + i \mu ) \simeq - \gamma \mu$, with
$\gamma = 0.5772156649015328606 = \lim_{n\to \infty} (\sum_{k=1}^n 1/k -\ln n)$
denoting Euler's constant~\cite{AS}.

To determine the frequency normalization factor ${\cal N}_{\omega,n}$
of the vector mode functions ${\bf g}_{\omega,n}(\mathbf{x}) = \nabla
\times \mathbf{G}_{\omega,n}(\mathbf{x})$, we start from the
differential identity
\begin{equation}
  ( \nabla \times \mathbf{A}  )  \cdot  (  \nabla \times \mathbf{A} ) =
  \nabla \cdot [ \mathbf{A} \times ( \nabla \times \mathbf{A} ) ] + 
  \mathbf{A} \cdot [ \nabla  ( \nabla \cdot \mathbf{A})  - \nabla^2 \mathbf{A}] ,
\end{equation}
which is valid for any nonsingular vector field
$\mathbf{A}(\mathbf{x})$.  Using Gauss theorem together with the fact
that the mode functions fulfill the relations $\nabla \cdot
\mathbf{G}_{\omega,n}(\mathbf{x}) = 0$ and
$\mathbf{G}_{\omega,n}(\mathbf{x}) \times \nabla \times
\mathbf{G}_{\omega,n}(\mathbf{x}) \mid_{\partial V} = 0$ on the
surface of the parabola $\partial V$, we obtain the normalization
condition
\begin{eqnarray}
  & \displaystyle
  \int_V d^3\mathbf{x}~
  [  \nabla \times  \mathbf{G}_{\omega,n}( \mathbf{x})^{\ast} ]  \cdot
  [ \nabla \times  \mathbf{G}_{\omega',n'}(\mathbf{x}) ]  =
  \frac{\omega'^2}{c^2} & 
 \nonumber  \\
  & \displaystyle
  \times   \int_V d^3\mathbf{x}~
  \mathbf{G}_{\omega,n}(\mathbf{x})^{\ast} \cdot
  \mathbf{G}_{\omega',n'} (\mathbf{x}) \, , &
  \label{integral1}
\end{eqnarray}
where $V$ denotes the volume bounded by the parabola.  The integral on
the right-hand side can be evaluated with the help of the
one-dimensional differential equations (\ref{Coulomb}). In this way,
we obtain the relation for $n = n^{\prime}$, for example,
\begin{eqnarray}
  &\frac{1}{\sqrt{\omega^{\prime} \omega \mathcal{N}_{\omega',n} \mathcal{N}_{\omega,n}}}
  W(\chi_{\omega^{\prime},n}(\xi), \chi_{\omega.n}(\xi)) 
  \int_0^{2f} d \eta~
  \frac{\chi_{\omega',n}({\eta}) \chi_{\omega,n}(\eta)}{\eta}
  & \nonumber \\
  & \displaystyle
  = (\omega' - \omega)\frac{\omega' + \omega}{c^2} 
  \int_V d^3\mathbf{x}~
  \mathbf{G}_{\omega',n}(\mathbf{x})^{\ast} \cdot
  \mathbf{G}_{\omega,n}({\bf x}) \, , & 
\end{eqnarray}
with the Wronskian-type quantity
\begin{eqnarray}
  W(\chi_{\omega',n}(\xi), \chi_{\omega.n}(\xi)) & = &
  \lim_{\xi \to \infty} \left [ 
    \chi_{\omega',n}(\xi)\frac{d\chi_{\omega,n}}{d\xi}(\xi) \right . 
  \nonumber \\
  & - &
  \left . \chi_{\omega,n}(\xi)\frac{d\chi_{\omega',n}}{d\xi}(\xi)
  \right  ] \, .
  \label{asymptoticnormalization}
\end{eqnarray}
 With the help of Eq.~(\ref{asymptoticnormalization}), the normalization constant ${\cal
  N}_{\omega,n}$ can be determined in such a way that the frequency
normalization condition of Eq.~(\ref{orthonormal}) is fulfilled.  For
this purpose we use the following representation of the Dirac
$\delta$ function
\begin{equation}
  \delta(\omega^{\prime} - \omega)  = \lim_{\xi \to \infty} 
\frac{\sin [(\omega^{\prime} - \omega)\xi]}
 {\pi(\omega^{\prime}-\omega)} \, .
\end{equation}
This allows us to find the explicit form of the
normalization factor given in  Eq.~(\ref{normal1}).


\begin{thebibliography}{99}
\bibitem{Berman:1994cd}
P. R. Berman (ed.), 
\emph{Cavity Quantum Electrodynamics} 
(Academic Press, San Diego, 1994).

\bibitem{Walther:2006qc}
H. Walther, B. T. H. Varcoe, B.-G. Englert, and T. Becker, 
Rep. Prog. Phys. \textbf{69}, 1325–(2006). 

\bibitem{Haroche:2006pm}
S. Haroche and J.-M. Raimond, 
\emph{Exploring the Quantum: Atoms, Cavities and Photons} 
(Oxford University Press, Oxford, 2006).

\bibitem{Wallraff:2004hj} 
A. Wallraff, D. I. Schuster, A. Blais, L. Frunzio, R.-S. Huang, 
J. Majer, S. Kumar, S. M. Girvin, and  R. J. Schoelkopf, 
Nature \textbf{431}, 162 (2004).

\bibitem{Colombe:2007qc}
Y. Colombe, T. Steinmetz, G. Dubois, F. Linke, D. Hunger, and J. Reichel,
Nature \textbf{450}, 272 (2007);
M. F. Riedel, P. Böhi, Yun Li, T. W. \"ansch, A. Sinatra, and P. Treutlein,
{\em ibid.} \textbf{464}, 1170 (2010).

\bibitem{Aoki:2006jk}
T. Aoki, B. Dayan, E. Wilcut, W. P. Bowen, A. S. Parkins, 
T. J. Kippenberg,  K. J. Vahala, and H. J. Kimble,
Nature  \textbf{443}, 671 (2006);
B. Dayan, A. S. Parkins, T. Aoki, E. P. Ostby,  K. J. Vahala,
H. J. Kimble, 
Science \textbf{319}, 1062 (2008).

\bibitem{Boundaries}
T. Boyer, 
Phys. Rev. \textbf{174}, 1764 (1968);
 K. Kakazu and Y. S. Kim, 
Phys. Rev. A \textbf{50}, 1830 (1994);
V. V. Klimov, V. S. Letokhov, and M. Ducloy, 
{\em ibid} \textbf{56},  2308 (1997);
J. U. N\"ockel, G. Bourdon, E. Le Ru, R. Adams,  I. Robert,
J.-M. Moison, and I. Abram, 
Phys. Rev. E  \textbf{62}, 8677 (2000);
R. Dubertrand, E. Bogomolny, N. Djellali, M. Lebental, and C. Schmit, 
Phys. Rev. A \textbf{77}, 013804 (2008);
F. S. S. Rosa, T. N. C. Mendes, A. Ten\'orio, and C. Farina,
{\em ibid}   \textbf{78}, 012105 (2008);
G. H\'etet, L. Slodi\v{c}ka, A. Gl\"{a}tzle, M. Hennrich, and R. Blatt, 
{\em ibid} \textbf{82}, 063812 (2010). 


\bibitem{radmodes} 
P. Goy, J. M. Raimond, M. Gross, and S. Haroche,
Phys. Rev. Lett.  \textbf{50}, 1903 (1983);
R. J. Cook and P. W. Milonni, 
Phys. Rev. A  \textbf{35}, 5081  (1987);
D. J. Heinzen, J. J. Childs, J. E. Thomas, and  M. S. Feld, 
Phys. Rev. Lett.  \textbf{58}, 1320 (1987);
F. De Martini, G. Innocenti, G. R. Jacobowitz, and P. Mataloni, 
{\em ibid}  \textbf{59}, 2955 (1987).

\bibitem{contmodes} 
I. Gerhardt, G. Wrigge, P. Bushev, G. Zumofen,   M. Agio, R. Pfab, and
V. Sandoghdar, 
Phys. Rev. Lett. \textbf{98}, 033601 (2007);
A. N. Vamivakas, M. Atat\"ure, J. Dreiser, S. T. Yilmaz, A. Badolato, 
A. K. Swan, B. B. Goldberg, A. Imamoglu, and M. S. \"{U}nl\"{u},
Nano Lett. \textbf{7}, 2892 (2007);
G. Zumofen, N. M. Mojarad, V. Sandoghdar, and   M. Agio, 
Phys. Rev. Lett. \textbf{101}, 180404 (2008);
 G. Wrigge, I. Gerhardt, J. Hwang, G. Zumofen, and  V. Sandoghdar, 
Nature Phys. \textbf{4}, 60 (2008);
M. K. Tey, Z. Chen, S. A. Aljunid, B. Chng, F. Huber, G. Maslennikov,
and Ch. Kurtsiefer, 
{\em ibid} \textbf{4}, 924 (2008);
L. Slodi\v{c}ka, G. H\'etet, S. Gerber, M. Hennrich, and R. Blatt, 
Phys. Rev. Lett. \textbf{105}, 153604 (2010); 
Y. Wang, J. Min\'{a}\v{r}, L. Sheridan, and V. Scarani,
Phys. Rev. A \textbf{83}, 063842 (2011).

\bibitem{focusing} 
S. J. van Enk and H. J. Kimble,  
Phys. Rev. A \textbf{63}, 023809 (2001);
H. P. Urbach and S. F. Pereira,
Phys. Rev. Lett. \textbf{100}, 123904 (2008).

\bibitem{matching}
I. M. Basset, 
J. Mod. Opt.\textbf{33}, 279 (1986);
S. J. van Enk, 
Phys. Rev. A \textbf{69}, 043813   (2004);
N. Bokor and N. Davidson, 
Opt. Lett. \textbf{29}, 1968 (2004);
M. Sondermann, R. Maiwald, H. Konermann,  N. Lindlein, U. Peschel and
G. Leuchs, 
Appl. Phys. B \textbf{89}, 489  (2007);
 P. Pinotsi and A. Imamoglu, 
Phys. Rev. Lett. \textbf{100},  093603 (2008).

\bibitem{tailorpol}
S. Quabis, R. Dorn, M. Eberler, O. Gl\"ockl, and  G. Leuchs, 
Opt. Commun. \textbf{179}, 1 (2000); 
R. Dorn, S. Quabis, and G. Leuchs, 
Phys. Rev. Lett. \textbf{91},  233901 (2003);
S. A. Aljunid, G. Maslennikov, Y. Wang, D. H. Lan, V. Scarani, and
Ch. Kurtsiefer, arXiv:1304.3761.

\bibitem{Alber:1992qr} 
G. Alber, 
Phys. Rev. A  \textbf{46}, R5338 (1992).

\bibitem{Viviescas:2003pi} 
C. Viviescas and G. Hackenbroich, 
Phys. Rev. A \textbf{67}, 013805 (2003).

\bibitem{Daul:2005se}
J.-M. Daul and P. Grangier,  
Eur. Phys. J. D \textbf{32}, 181(2005). 

\bibitem{Dorner:2002pz} 
U. Dorner and P. Zoller, 
Phys. Rev. A \textbf{66}, 023816   (2002). 

\bibitem{StAl}  M. Stobi\'nska and R. Alicki, Open Syst. Inf. Dyn. {\bf 19}, 1250023 (2012).

\bibitem{Eschner:2001aj} 
J. Eschner, Ch. Raab, F. Schmidt-Kaler, and  R. Blatt, 
Nature \textbf{413}, 495 (2001).

\bibitem{StSoLe}  M. Stobi\'nska, M. Sondermann, and G. Leuchs, Opt. Commun., {\bf 283}, 737 (2010).

\bibitem{Bokor:2008pi}
N. Bokor and N. Davidson,
Opt. Commun. \textbf{281}, 5499 (2008).

\bibitem{Maiwald:2012ex}
R. Maiwald, A. Golla, M. Fischer, M. Bader, S. Heugel, B. Chalopin,
M. Sondermann, and G. Leuchs,
Phys. Rev. A\textbf{86}, 043431 (2012). 

\bibitem{AS} 
M. Abramowitz and I. A. Stegun, eds.,  
\emph{Handbook of Mathematical Functions} (Dover,  New York, 1972).

\bibitem{Boyer:1976vt} 
C. P. Boyer, E. G. Kalnins, and W. Miller, 
Nagoya  Math. J. \textbf{60}, 35 (1976).

\bibitem{Feynman} 
L.S. Schulman, 
{\em Techniques and Applications of Path Integration} (Wiley, New
York, 1981).

\bibitem{poleapproximation} 
 V. Weisskopf and E. Wigner, 
Z. Phys. \textbf{63}, 54   (1930).

\bibitem{GoldenRule} 
C. Cohen-Tannoudji, B. Diu, and F. Laloe, 
{\em Quantum Mechanics} (Wiley, New York, 1977).

\bibitem{PoissonSummation} 
P. M. Morse and H. Feshbach, {\it Methods of Theoretical Physics}, 
(McGraw, New York, 1953).

\bibitem{Maslov} 
V. P. Maslov and M. V. Fedoriuk, 
{\em Semi-Classical Approximation in Quantum Mechanics} 
(Reidel, Dordrecht, 1981).

\bibitem{LambShift1} 
H. A. Bethe, 
Phys. Rev. {\bf 72}, 339 (1947);
J. Seke, 
Nuovo Cimento D {\bf 18}, 533 (1996).

\bibitem{expodecaydeviation} 
S. Geltman,  J. Phys. B {\bf 10}, 831 (1977). 

\bibitem{LeSo} G. Leuchs and M. Sondermann, J. Mod. Opt. {\bf 60}, 36 (2013).

\bibitem{StoAlLe} M. Stobi\'nska, G. Alber, and G. Leuchs, EuroPhys. Lett. {\bf 86}, 14007 (2009).

\bibitem{Glauber} 
R. Glauber, 
Phys. Rev. {\bf 130}, 2529 (1963); \emph{ibid.}  {\bf 131}, 2766
(1963);
P. L. Kelly and W. H. Kleiner, 
{\em ibid}  {\bf 136}, 316 (1964); 
H. Carmichael, 
{\em An Open System Approach to Quantum Optics},
Lect. Notes Phys. {\bf 18} (Springer, Berlin, 1991).

\bibitem{34} A. Golla, B. Chalopin, M. Bader, 
I. Harder, K. Mantel, R. Maiwald, N. Lindlein, M. Sondermann, and G. Leuchs, Eur. Phys. J. D {\bf 66}, 190 (2012). 

\end{thebibliography}
\end{document}